\documentclass[twocolumn,groupedaddress,showpacs,nofootinbib]{revtex4-1}

\usepackage[colorlinks,pdfstartview=FitH]{hyperref}
\hypersetup{linkcolor=blue,citecolor=blue,filecolor=black,urlcolor=blue}

\usepackage{amsmath,amssymb,bm}
\usepackage{slashed}
\usepackage{graphicx}
\usepackage{amsfonts}
\usepackage{color}

\usepackage{amsfonts}
\usepackage{bbm}
\usepackage{latexsym}

\begin{document}

\title{Study of neutrinophilic low-mass dark matter mediated by pseudoscalar}

\author{Zhuo Zhang}
\affiliation{
School of Mathematics and Physics, Southwest University of Science and Technology, Mianyang 621010, China}

\author{Lian-Bao Jia}  \email{jialb@mail.nankai.edu.cn}
\affiliation{
School of Mathematics and Physics, Southwest University of Science and Technology, Mianyang 621010, China}

\author{Reyes J.F. Eduardo}
\affiliation{
School of Mathematics and Physics, Southwest University of Science and Technology, Mianyang 621010, China}

\begin{abstract}

In this work, we investigate a neutrinophilic low-mass dark matter model mediated by a pseudoscalar particle. Since dark matter lacks Standard Model gauge charges, new interactions are required to connect it to the visible sector. Traditional indirect detection searches for annihilation products, such as cosmic rays, become ineffective when the annihilation predominantly yields invisible neutrinos. In our model, the present-day annihilation cross section into neutrinos (manifesting as a neutrino line) falls below current indirect detection limits. We therefore constrain the model using complementary probes: the Lyman-$\alpha$ forest, high-energy astrophysical neutrinos from active galactic nuclei and supernovae, direct detection via nucleon and electron scattering, and invisible Higgs decays. These observables provide stringent and multifaceted constraints on neutrinophilic dark matter interactions in the low-mass regime. Our results indicate that searches for the neutrino line from dark matter annihilations, neutrino self-interactions from supernovae and collider signatures, and invisible Higgs decays offer critical tests for the model's parameter space.

\end{abstract}


\maketitle

\section{Introduction}

The existence of dark matter (DM) was found many decades ago, yet the nature of DM is still unclear. Assuming DM was in thermal equilibrium with the standard model (SM) particles at high temperature in the early Universe, the observed relic DM is produced via the thermal freeze-out mechanism. For such thermal freeze-out DM, the allowed mass range spans from MeV to TeV \cite{Griest:1989wd}. In particular, the mass of DM is typically constrained to be $\gtrsim$ 10 MeV by big bang nucleosynthesis (BBN) and cosmic microwave background (CMB) measurements \cite{Ho:2012ug,Boehm:2013jpa,Wilkinson:2016gsy,Jia:2016uxs,Berlin:2018sjs}. Since DM does not carry SM gauge charges, new interactions are needed to bridge it to the SM sector, and the possible new physics signatures may be disclosed in precise measurements, such as searches for excesses of cosmic rays (gamma rays, positrons, etc.) from DM annihilations. On the SM side, another possible portal is through new interactions with neutrinos \cite{Fiorillo:2020jvy,Du:2020avz,Berryman:2022hds,Bell:2024uah,Dev:2025tdv}. Here we focus on the scenario where the annihilation into neutrinos is dominant in DM annihilations. In this case, the corresponding new physics might be hidden in the invisible products, and the test of such scenarios becomes challenging.

For neutrinophilic DM, the exploration of its interactions with visible matter via neutrinos is a challenge, and the monochromatic neutrino line from DM annihilations serves as a potential signature in indirect detections \cite{Arguelles:2019ouk,Akita:2022lit,IceCube:2023ies}. Considering a new pseudoscalar $\phi$ as the mediator in the annihilations of fermionic DM $\chi$, which can have a coupling with the Higgs field. When $m_{\phi}$ is much smaller than the electroweak scale, its loop contribution to the Higgs mass is negligible. Besides the s-wave annihilation of DM, the p-wave process $\chi \bar{\chi} \to \phi \phi$ opens when $m_{\phi} \leq m_{\chi}$, and becomes important for the DM freeze-out process in the early universe. In this case, the present-day annihilation cross section into neutrinos can fall below the canonical value of $(2-4)\times 10^{-26}$ cm$^3$/s associated with thermal freeze-out. For direct detection, neutrinophilic DM scattering off nuclei can occur at the one-loop level, which in principle offers a complementary probe. Meanwhile, the recoil energy of low-mass DM ($m_{\chi} < 10$ GeV) is intrinsically smaller than that of the higher-mass region ($m_{\chi} \gtrsim 10$ GeV), making the low-mass range a particularly interesting target in DM searches. In addition, neutrinophilic dark matter can contribute to invisible final states in collider searches. This type DM will be explored in this paper.

\section{Interactions and annihilations}

In this section, we consider Dirac fermionic DM $\chi$ and a leptophilic pseudoscalar $\phi$ in a new sector. The effective interactions of $\phi$ with DM, neutrinos and Higgs field $H$ are written as
\begin{eqnarray}
\mathcal{L}_\mathrm{int} &\supset& - \phi \bar \chi (  i\lambda_p \gamma^5) \chi - \frac{1}{2}\phi \bar \nu_\alpha ( ig_p^{\alpha\beta} \gamma^5) \nu_\beta  \nonumber \\
&& - \lambda_h \phi^2 H^\dag H  \, ,
\end{eqnarray}
where $\lambda_p$ and $\lambda_h$ are couplings parameters. $g_p^{\alpha\beta}$ is the effective coupling of $\phi$ to Majorana neutrinos in the flavor basis with $\alpha$, $\beta$ = $e$, $\nu$, $\tau$. The interaction is assumed to be diagonal and universal for simplicity, i.e., the only non-zero entries are $g_p^{\alpha\alpha}$ = $g_p$. In addition, we should remember that there may be more particles in the new sector, and only particles play important roles in the transitions between DM sector and SM sector are considered here.

\begin{figure}[htbp!]
\includegraphics[width=0.45\textwidth]{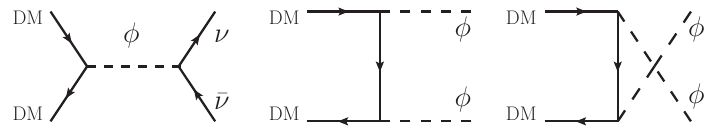} \vspace*{-1ex}
\caption{The annihilation process of DM.}
\label{neu1}
\end{figure}

The annihilation modes of DM are related to the mass relation between $m_\chi$ and $m_\phi$. Here we consider the case of $2 m_{\phi} < 2 m_{\chi} < $ 3$m_\phi$, with DM cascade annihilations allowed in the early universe and a sizable neutrino line in the annihilation products today.\footnote{The $s-$wave channel 2$m_\chi \to$ 3$m_\phi$ will be opened when 2$m_\chi > $ 3$m_\phi$, see e.g. Ref. \cite{Jia:2018mkc}.} The annihilation process is shown in Fig. \ref{neu1}. The annihilation cross sections of the $s-$wave process $\bar{\chi}\chi \to \bar{\nu} \nu$ and $p-$wave process $\bar{\chi}\chi \to \phi \phi$ are
\begin{eqnarray}
\sigma_1 v_r \simeq \frac{1}{2} \frac{\lambda_p^2  g_p^2 }{32 \pi (s - 2 m_\chi^2)} \frac{s^2}{(s-m_\phi^2 )^2 } ,
\end{eqnarray}
\begin{eqnarray}
\sigma_2 v_r \simeq \frac{1}{2} \frac{\lambda_p^4 ( 1 - 4 m_\phi^2 /s)^\frac{5}{2} }{48 \pi (s - 2 m_\chi^2)} \frac{ s^3(s - 4 m_\chi^2 )}{(s- 2 m_\phi^2)^4 } ,
\end{eqnarray}
respectively. Here $v_r$ is the relative velocity, the factor $\frac{1}{2}$ is for the $\bar{\chi}\chi$ pair required in DM annihilations, and $s$ is the total invariant mass squared. The annihilation mediated by the Higgs boson $h$ occurs at one-loop level and is suppressed. Thus, the total annihilation cross section is $\sigma_{ann} v_r \simeq \sigma_1 v_r +\sigma_2 v_r$.

\section{Analysis and test}

Here we give an analysis about the new sector with the corresponding constraints. Some parameters
are inputted as follows: $m_h^{}$ = 125 GeV \cite{ATLAS:2015yey,ParticleDataGroup:2024cfk}, $m_N^{} \simeq (m_p^{} + m_n^{})/2$ = 0.939 GeV \cite{ParticleDataGroup:2024cfk}, $m_e$ = 0.511 MeV  and $m_{\chi}$ in a range between 10 MeV and 5 GeV.

\subsection{Relic density, indirect detection and neutrino self-interaction}

\begin{figure}[htbp!]
\includegraphics[width=0.42\textwidth]{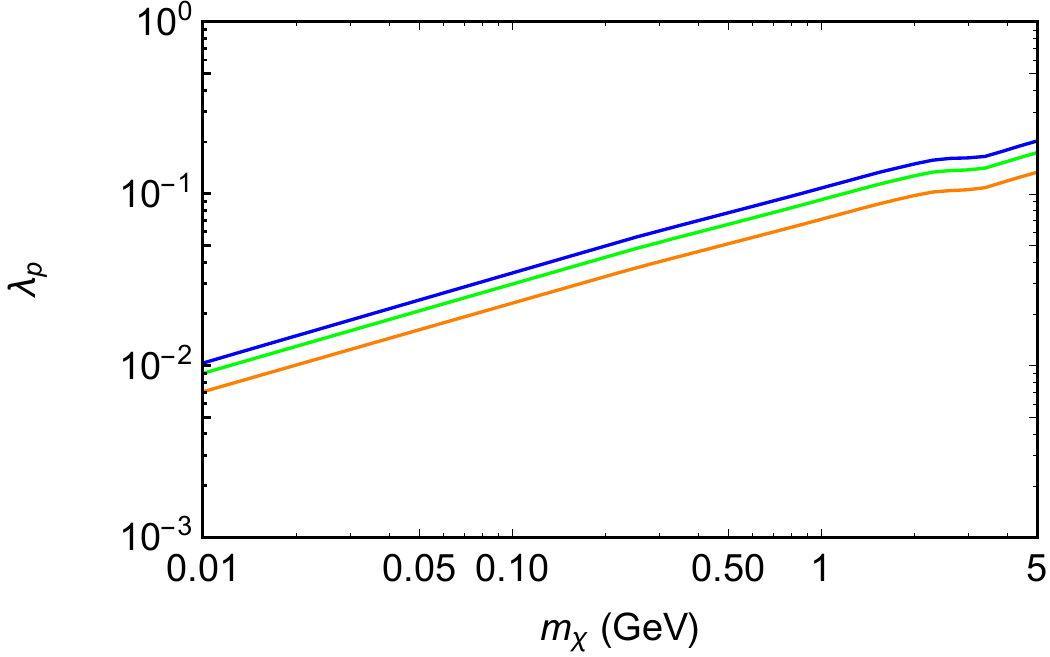} \vspace*{-1ex}
\caption{The couplings parameter $\lambda_p$ as a function of DM mass $m_{\chi}$ with $m_{\phi} = 0.8 m_{\chi}$ adopted. The solid curves from top to bottom are for the cases of $k =$ 0.5, 1, 2, respectively.}
\label{lp}
\end{figure}

\begin{figure}[htbp!]
\includegraphics[width=0.42\textwidth]{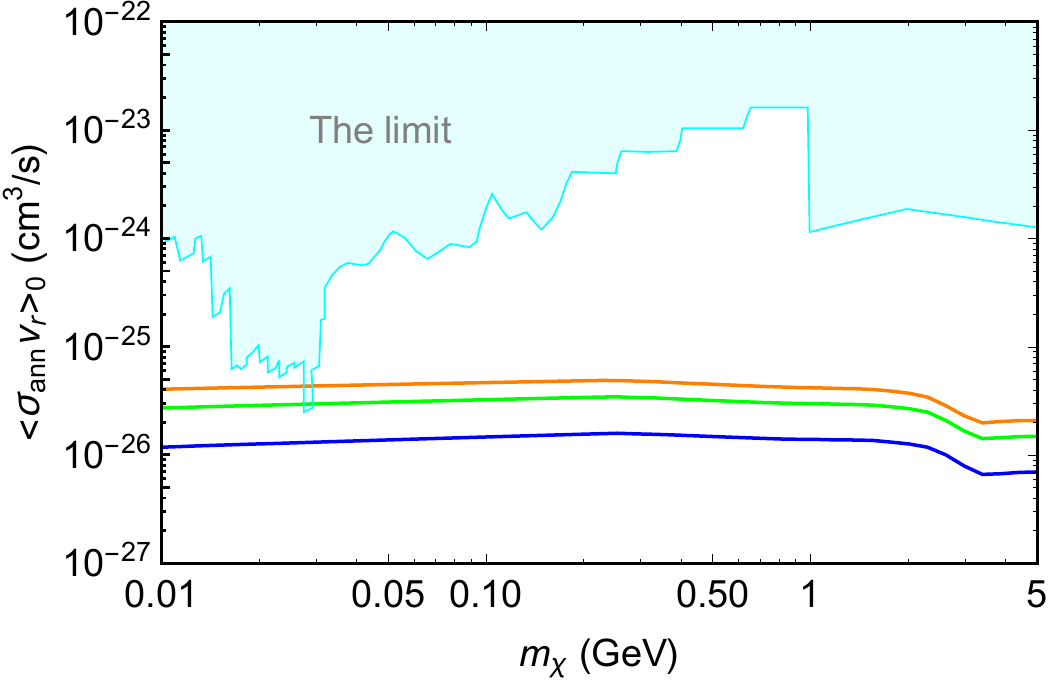} \vspace*{-1ex}
\caption{The DM annihilation cross section into neutrinos $\langle\sigma_{ann} v_r\rangle_0$ as a function of DM mass $m_{\chi}$ with $m_{\phi} = 0.8 m_{\chi}$ adopted. The solid curves from top to bottom are for the cases of $k =$ 2, 1, 0.5, respectively. The shaded region is excluded by experimental limits \cite{Arguelles:2019ouk}.}
\label{anncs}
\end{figure}

The relic density of DM is related to the annihilation cross section, and the coupling parameters can be set by this constraint. The DM relic density today is 0.120$\pm$0.001 \cite{Planck:2018vyg}. Taking a mass relation of $m_{\phi} \sim 0.8 m_{\chi}$, the coupling parameter $\lambda_p$ between DM and $\phi$ can be derived with a typical value $g_p/\lambda_p = k\times 10^{-1}$ adopted, as shown in Fig. \ref{lp}. In addition, the DM annihilation cross section into neutrinos today $\langle\sigma_{\mathrm{ann}} v_r\rangle_0$ can be derived, mostly contributed by $\sigma_1 v_r$, and the annihilation yields a monochromatic neutrino line. The result is shown in Fig.~\ref{anncs} alongside the experimental limits \cite{Arguelles:2019ouk}. The neutrino line from DM annihilation process may be detectable by next-generation neutrino experiments.

Additionally, in the pseudoscalar-mediated DM scenario, the neutrino-DM scattering cross section exhibits a strong energy dependence,
\begin{eqnarray}
\sigma_{\nu\chi} = \frac{\lambda_p^2 g_p^2}{16 \pi s} \int \frac{d \cos \theta}{2} \frac{q^4}{(q^2 - m_\phi^2)^2},
\end{eqnarray}
where $q$ is the momentum transfer. For incident neutrino energies much smaller than the DM mass, $E_\nu \ll m_\chi$, the cross section is deeply suppressed due to the $q^4$ scaling. In particular, for the cosmic neutrino background with $E_\nu \lesssim 10^{-4}\,\mathrm{eV}$, this suppression renders Lyman-$\alpha$ forest constraints ineffective \cite{Wilkinson:2014ksa,Hooper:2021rjc}. For high-energy neutrinos from active galactic nuclei (AGN), the dense DM spike surrounding the central supermassive black hole in AGNs. The cross section scales as $\sigma_{\nu\chi} \propto 1/E_{\nu}$ here. Existing bounds derived under the assumption of a constant cross section or a cross section that scales linearly with energy \cite{Cline:2023tkp} are not directly applicable to the scenario here due to the distinct energy scaling. As a rough estimate, for DM masses from $10\,\mathrm{MeV}\) to \(1\,\mathrm{GeV}$, $\sigma_{\nu\chi}$ varies from $10^{-28}\,\mathrm{cm}^2$ to $10^{-27}\,\mathrm{cm}^2$. This implies that the predicted scattering cross sections lie approximately $10^{-13}\) to \(10^{-11}$ orders of magnitude below current constraints.

For a light $\phi$ particle, it can be radiated through its coupling to neutrinos, thereby affecting the cooling of core-collapse supernovae. The impact of such a light pseudoscalar on the SN1987A observations was studied in Ref.~\cite{Heurtier:2016otg}, and the analysis shows that for $1\,\mathrm{MeV} \lesssim m_\phi \lesssim 100\,\mathrm{MeV}$, coupling values in the range $10^{-11} \lesssim g_{p} \lesssim 10^{-6}$ are excluded. For larger couplings, $\phi$ particles become trapped and reabsorbed by the supernova, thus no longer affecting its cooling. Consequently, this bound corresponds to a relatively weak coupling regime that is much smaller than the typical value considered in this paper.

\begin{figure}[htbp!]
\includegraphics[width=0.42\textwidth]{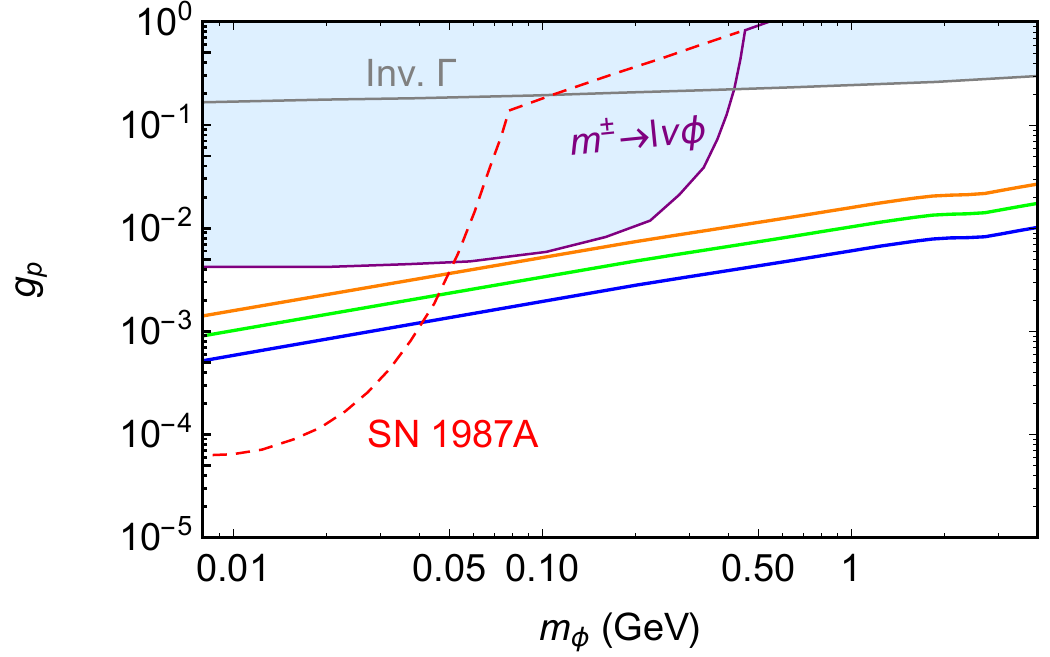} \vspace*{-1ex}
\caption{The coupling parameter $g_p$ as a function of $m_{\phi}$ with $m_{\phi} = 0.8 m_{\chi}$ adopted. The solid curves from top to bottom are for the cases of $k =$ 2, 1, 0.5, respectively. The shaded region is laboratory constraints \cite{Berryman:2022hds}, and the dashed curve represents the bound from the SN 1987A burst outflow \cite{Chang:2022aas}.}
\label{neu-s}
\end{figure}

Here we turn to neutrino self-interactions. Neutrino self-interactions are extremely weak in the SM, but they may be enhanced by new interactions (here is the new interaction mediated by $\phi$). Such interactions could have implications for cosmology, astrophysics, and laboratory experiments \cite{Berryman:2022hds}. Indeed, supernovae are expected to be powerful probes for investigating neutrino self-interactions, as the high core densities ensure that even feeble self-interaction leads to frequent scattering. The s-channel $\bar{\nu}\nu \to \phi \to \bar{\nu}\nu$ with resonant effect is dominant for a MeV mediator, and the cross section is
\begin{eqnarray}
\sigma_{\bar{\nu}\nu} \simeq \frac{g_p^4}{32 \pi}\frac{s}{ (s - m_\phi^2)^2+m_\phi^2 \Gamma_\phi^2} ,
\end{eqnarray}
with the width of $\phi$ being $\Gamma_\phi= g_p^2 m_\phi /16 \pi$. Recent work \cite{Chang:2022aas} has shown that such interactions can significantly alter the neutrino outflow from a protoneutron star. In the burst-outflow scenario, which yields the strongest sensitivity, the expanding neutrino ball remains homogeneous until decoupling, and the observed signal duration is directly linked to the interaction strength. The SN1987A data translate into an upper limit on the coupling $g_p \lesssim 10^{-4}$ for a mediator mass $m_\phi \sim 10$ MeV \cite{Chang:2022aas}, and the constraints on $g_p$ are shown in Fig.~\ref{neu-s}, along with laboratory constraints \cite{Berryman:2022hds}. For the typical values of $g_p$, the mass $m_\phi \gtrsim$ 40 MeV is allowed by constraints. Future supernova analyses are poised to probe couplings down to $10^{-5}$, thus establishing supernovae as a powerful and complementary probe of neutrino self-interactions alongside future collider experiments \cite{Liu:2024ywd}.

\subsection{Direct detection and invisible decay}

\begin{figure}[htbp!]
\includegraphics[width=0.19\textwidth]{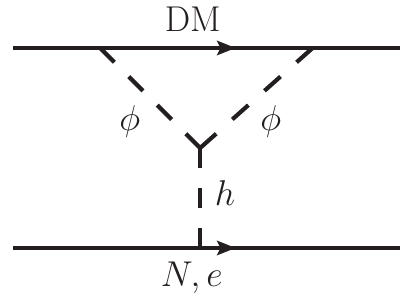} \vspace*{-1ex}
\caption{The scattering in DM direct detections.}
\label{dmdd}
\end{figure}

In low-mass DM direct detections, the scattering between DM particle $\chi$ and target material (nucleon $N$ and electron $e$) occurs at one-loop level, as shown in Fig. \ref{dmdd}. At the zero momentum transfer limit, the spin-independent scattering cross section between $\chi$ and $N$ is
\begin{eqnarray}
\sigma_{N}^{} = \frac{4 \lambda_p^4 \lambda_h^2 g_{hNN}^2 v^2 m_\chi^2 m_N^2 c_l^2}{\pi (m_\chi +m_N)^2  m_h^4}  ,
\end{eqnarray}
where $v$ is the vacuum expectation value in SM. Here $g_{hNN}^{}$ is the effective Higgs-nucleon coupling parameter, with $g_{hNN}^{} \approx 1.1 \times 10^{-3}$ \cite{Cheng:2012qr,Ellis:2000ds,Gondolo:2004sc,He:2008qm,Alarcon:2011zs,Cline:2013gha} adopted in calculations. The loop parameter $c_l^{}$ is
\begin{eqnarray}
c_l^{} = \frac{1}{16 \pi^2}\int_0^1 dx \frac{x (1-x) m_\chi^{}}{(1-x) m_\phi^2 +x^2m_\chi^2}  .
\end{eqnarray}
To the scattering between $\chi$ and $e$, the cross section is
\begin{eqnarray}
\bar{\sigma}_{e}^{} = \frac{4 \lambda_p^4 \lambda_h^2 m_e^4 m_\chi^2 c_l^2}{\pi (m_\chi +m_e)^2  m_h^4}  ,
\end{eqnarray}
with the form factor $F_{\mathrm{DM}} (q) = 1$ \cite{Essig:2011nj}.

Since the parameter $\lambda_h$ needs to be determined prior to the numerical analysis of DM direct detection, we estimate it using the invisible decay of the Higgs boson. The invisible Higgs decay $h \to \phi \phi$ followed by $\phi \to$ neutrinos may be sizable, which can be examined at future Higgs factories. To the invisible channel $h \to \phi \phi$ with $\phi \to \bar{\nu} \nu$, the decay width $\Gamma_{\phi\phi}$ is
\begin{eqnarray}
\Gamma_{\phi\phi} = \frac{\lambda_h^2 v^2}{8 \pi m_h^{}} \sqrt{1-\frac{4 m_{\phi}^2}{m_h^2} } ,
\end{eqnarray}
In addition, the DM pair decay $h \to \bar{\chi}\chi$ is one-loop suppressed, and the channel $h \to \phi^* \phi$ followed by $\phi^* \to \bar{\chi}\chi$ is also suppressed. The current upper limit on the invisible decay branching ratio of the Higgs boson $Br(\mathrm{Higgs}\to \mathrm{inv})$ is approximately 0.1 \cite{ATLAS:2023tkt,CMS:2023sdw}. In addition, the theoretical prediction of the total decay width $\Gamma_H$ for a SM Higgs boson with a mass of 125 GeV is 4.1 MeV with a relative uncertainty of $\pm 3.9\%$ \cite{LHCHiggsCrossSectionWorkingGroup:2016ypw}, which is consistent with the experimental measurements \cite{ATLAS:2024jry,CMS:2026igg}. Hence, we consider the case that the invisible width satisfies $\Gamma_{\phi\phi} \lesssim 0.1$ MeV, together with $\lambda_h \lesssim m_\phi / v$ in the calculation. In fact, invisible Higgs decays provide a complementary probe for future experiments, and future Higgs factories are expected to improve the sensitivity to such decays, potentially reaching a branching ratio of $\sim 0.002$ \cite{Ishikawa:2019uda,deBlas:2019rxi,Tan:2020ufz}.

\begin{figure}[htbp!]
\includegraphics[width=0.44\textwidth]{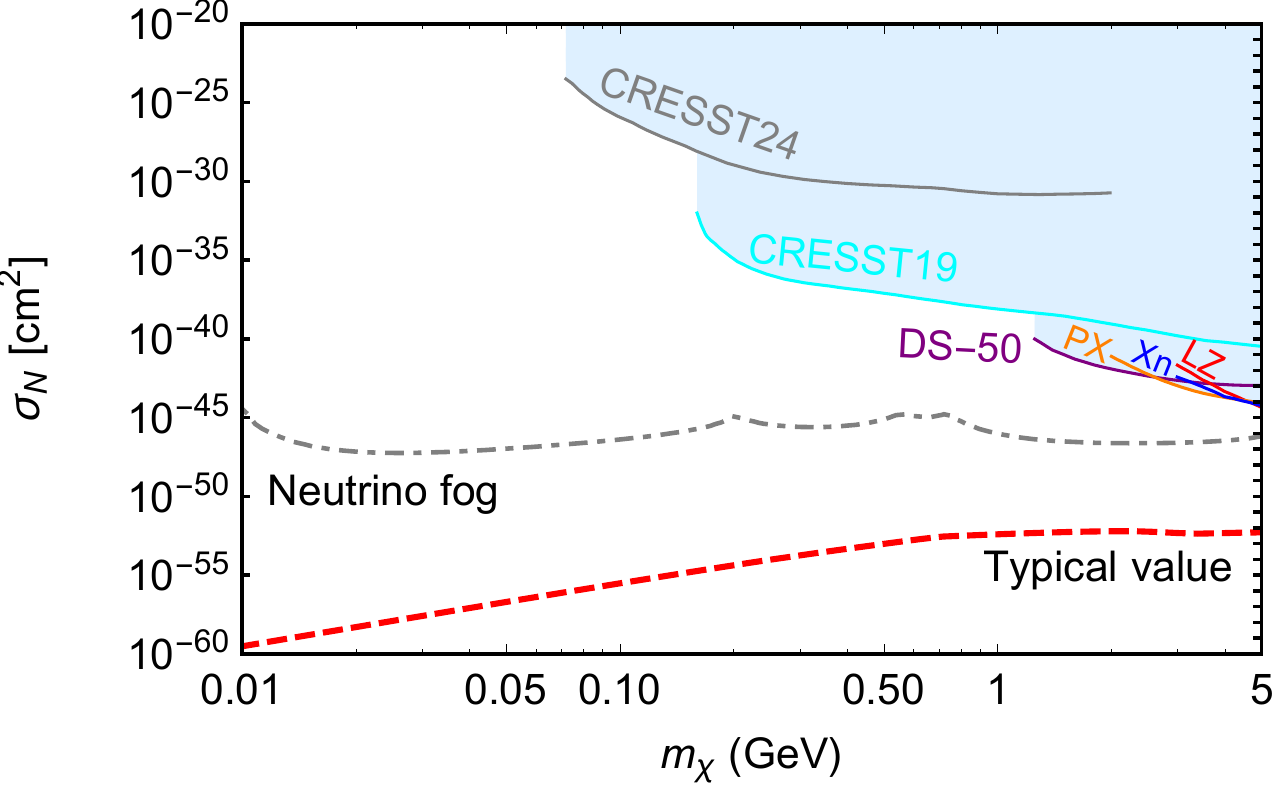} \vspace*{-1ex}
\caption{The DM-nucleon scattering cross section $\sigma_{N}$ as a function of DM mass $m_{\chi}$ with $m_{\phi} = 0.8 m_{\chi}$ adopted. The dashed curve corresponds to a typical value of $\sigma_{N}$ with $k = 1$. The shaded region denotes DM direct detection limits from DarkSide-50 (NQ) \cite{DarkSide-50:2025lns}, CRESST 2019 \cite{CRESST:2019jnq}, CRESST 2024 \cite{CRESST:2024cpr}, PandaX (PX) \cite{PandaX:2025rrz}, LZ \cite{LZ:2025igz}, and XENONnT (Xn) \cite{XENON:2026qow}. The dot-dashed curve represents the neutrino fog for a xenon target with $10^3$ ton$\cdot$year exposure \cite{AristizabalSierra:2021kht}.}
\label{dm-n}
\end{figure}

\begin{figure}[htbp!]
\includegraphics[width=0.44\textwidth]{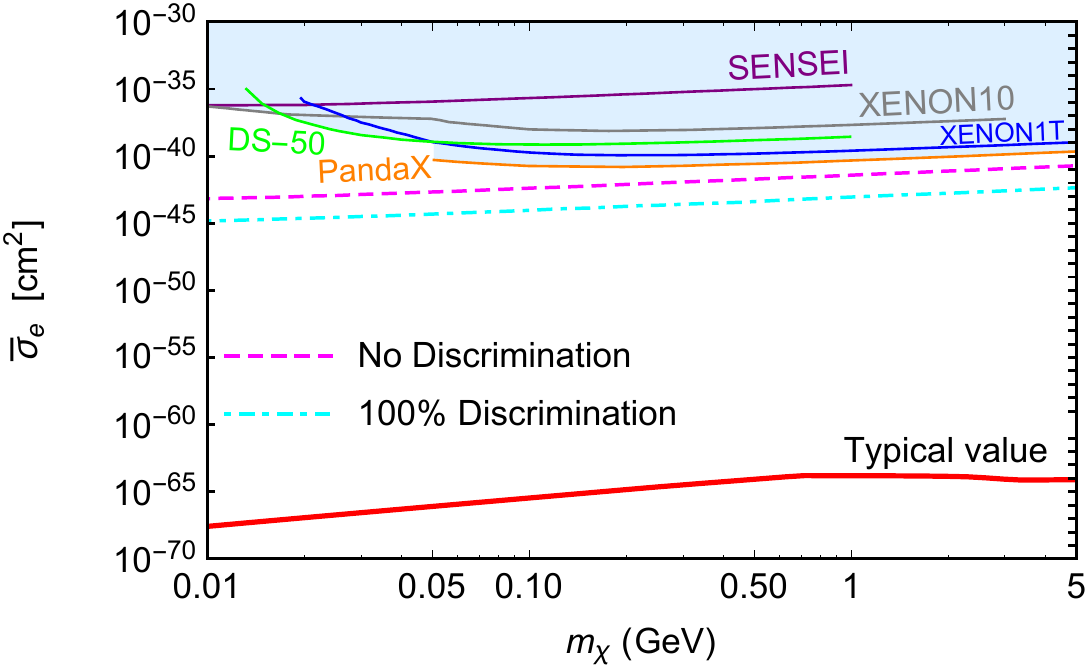} \vspace*{-1ex}
\caption{The DM-electron scattering cross section $\bar{\sigma}_{e}$ as a function of DM mass $m_{\chi}$ with $m_{\phi} = 0.8 m_{\chi}$ adopted. The solid curve corresponds to a typical value of $\bar{\sigma}_{e}$ with $k = 1$. The shaded region denotes DM direct detection limits from PandaX \cite{PandaX:2025rrz}, SENSEI \cite{SENSEI:2020dpa}, XENON10 \cite{Essig:2017kqs}, XENON1T \cite{XENON:2019gfn}, and DarkSide-50 \cite{DarkSide:2022knj}. The dashed and dot-dashed curves represent the neutrino backgrounds for an exposure of $10^5$ kg-yr \cite{Wyenberg:2018eyv}, with no discrimination and 100\% discrimination between electronic and nuclear scattering events, respectively.}
\label{dm-e}
\end{figure}

For the DM-nucleon scattering cross section, the result of $\sigma_{N}^{}$ is shown in Fig. \ref{dm-n}, alongside the limits from DM direct detections \cite{DarkSide-50:2025lns,CRESST:2019jnq,CRESST:2024cpr,PandaX:2025rrz,LZ:2025igz,XENON:2026qow}. Due to the loop-level suppression, a typical value of $\sigma_{N}^{}$ ($k = 1$ adopted) is much below the neutrino fog \cite{AristizabalSierra:2021kht}. Additionally, to the DM-electron scattering, the cross section $\bar{\sigma}_{e}^{}$ is shown in Fig. \ref{dm-e}, alongside the limits from DM direct detections \cite{PandaX:2025rrz,SENSEI:2020dpa,Essig:2017kqs,XENON:2019gfn,DarkSide:2022knj}. For DM-electron scattering, the discovery reach of direct detection experiments is limited by the neutrino background, with the discovery limits depending on the exposure and the ability to discriminate between electronic and nuclear recoils (a capability that traditional methods lack at low energies). The typical value of $\bar{\sigma}_{e}^{}$ ($k = 1$ adopted) is much below the neutrino background reported in Ref. \cite{Wyenberg:2018eyv} for an exposure of $10^5$ kg-yr, both with no discrimination and with 100\% discrimination between electronic and nuclear scattering events. Hence, it means that it is challenging to probe the DM of concern via the DM-nucleon and DM-electron scatterings.

Given the aforementioned challenges and limitations in DM direct detection, we instead consider alternative probes. For DM in the MeV-GeV range that annihilates into monoenergetic neutrinos via an s-channel process, the resulting flux appears as a narrow spectral line. Since astrophysical backgrounds are continuous, achieving high energy resolution in dedicated neutrino observatories such as JUNO \cite{JUNO:2021vlw},  Hyper-Kamiokande \cite{Hyper-Kamiokande:2018ofw} and DUNE \cite{DUNE:2024wvj} is essential for resolving this line and enabling a clean signature. In addition, DM direct detection experiments can also probe these monoenergetic neutrinos through coherent elastic neutrino-nucleus or neutrino-electron scattering, thereby offering a complementary approach that leverages their large target masses and low backgrounds.

\section{Conclusion and Discussion}
\label{sec:Con}

In this paper, we have investigated neutrino-DM interactions. We considered a neutrinophilic low-mass DM scenario in which a fermionic DM particle annihilates into neutrinos via a light pseudoscalar mediator, with the relic abundance determined by thermal freeze-out. The present-day annihilation cross section falls below current indirect detection limits. Hence, we examined various probes, including the Lyman-$\alpha$ forest, high-energy astrophysical neutrinos from active galactic nuclei, and direct detections via DM-nucleon and DM-electron scatterings. For the parameter space of interest, these probes do not provide meaningful constraints. In contrast, the neutrino line from DM annihilations, neutrino self-interactions in supernovae, collider signatures, and invisible Higgs decay are promising channels to test this DM model. In particular, for the monoenergetic neutrino line in the range $\gtrsim 10,\mathrm{MeV}$, the cross section can reach values accessible to neutrino experiments such as JUNO, Hyper-Kamiokande and DUNE, offering a viable probe. Taken together, these effective approaches (neutrino lines, supernova neutrinos, collider signatures, and invisible Higgs decay) provide a multifaceted strategy for constraining neutrino-DM interactions in the low-mass regime. Future improvements in experimental sensitivity will be essential to further test the viability of these interactions and to potentially uncover possible evidence of dark sector couplings.

\acknowledgments

This work was partly supported by Sichuan Science and Technology Program No. 2026NSFSC0758.


\end{document}